\documentclass[aps,showpacs,preprintnumbers,nofootinbib]{revtex4}
\usepackage{amsfonts}
\usepackage{amssymb}
\usepackage{amsmath}
\usepackage{bm}

\input{tcilatex}

\begin{document}

\title{Fermion confinement induced by geometry}
\author{$^{1}$ C. Romero\thanks{%
E-mail address: jemadriz@fisica.ufpb.br}, $^{1}$ J. B. Formiga\thanks{%
E-mail address: cromero@pq.cnpq.br} \ and \ $^{2}$ C. Dariescu \thanks{%
jfonseca@fisica.ufpb.br}. }
\affiliation{$^{1}$ Departamento de F\'{\i}sica, Universidade Federal da Para\'{\i}ba,
Caixa Postal 5008, 58059-970 Jo\~{a}o Pessoa, PB, Brazil\\
$^{2}$ Departament of Solid State and Theoretical Physics, Faculty of
Physics, "Al. I. Cuza University, Bd. Carol I no. 11, 700506 Iasi, Romania \\
E-mail: cromero@fisica.ufpb.br}

\begin{abstract}
We consider a five-dimensional model in which fermions are confined in a
hypersurface due to an interaction with a purely geometric field. Inspired
by the Rubakov-Shaposhnikov field-theoretical model, in which massless
fermions can be localized in a domain wall through the interaction of a
scalar field, we show that particle confinement may also take place if we
endow the five-dimensional bulk with a Weyl integrable geometric structure,
or if we assume the existence of a torsion field acting in the bulk. In this
picture, the kind of interaction considered in the Rubakov-Shaposhnikov
model is replaced by the interaction of fermions with a geometric field,
namely a Weyl scalar field or a torsion field. We show that in both cases
the confinement is independent of the energy and the mass of the fermionic
particle. We generalize these results to the case in which the bulk is an
arbitrary n-dimensional curved space.
\end{abstract}

\maketitle


\vskip .5cm


\section{Introduction}

It has been suggested in recent times that our world may be viewed as a
hypersurface (\textit{brane}) $\Sigma $ embedded in a higher-dimensional
manifold $M$, often referred to as \textit{the bulk }\cite{Roy}. As far as
the geometry of this hypersurface is concerned, it has been generally
assumed that it has a Riemannian geometrical structure. This hypothesis has
the great advantage of avoiding possible conflicts with the well-established
theory of general relativity which is formulated in a Riemannian geometrical
setting. Likewise there has not been much discussion\ on what kind of
geometry the bulk possesses, which is generally supposed to be also
Riemannian. Some attempts to broaden this scenario have appeared recently in
the literature, where non-Riemannian frameworks, in particular, Weyl and
Riemann-Cartan geometries, are taken into consideration as viable
possibilities to describe the geometry of the bulk \cite%
{Israelit,Arias,Nandinii,Shankar}.

In some brane-world scenarios it is postulated that the particles and fields
of the standard model are confined to the brane universe. Classically, test
particles are confined in spacetime if the hypersurface $\Sigma $ has
vanishing extrinsic curvature \cite{Seahra}. This kind of confinement is
purely geometrical, which means that the only force acting on the particles
is the gravitational force. Classical confinement has been investigated
recently in different contexts \cite{Seahra,Dahia}. At the quantum level,
the stability of the confinement of matter fields is made possible by
assuming an interaction of matter with a scalar field. An example of how
this mechanism works is clearly illustrated \ by a field-theoretical model
devised by Rubakov and Shaposhnikov, in which fermions may be trapped to a
brane by interacting with a scalar field that depends only on the extra
dimension \cite{Rubakov}. On the other hand, it has been shown that in a
purely classical (non-quantum) picture, there are geometrical mechanisms
which play the role of a quantum scalar field so as to constrain massive
particles to move on hypersurfaces in a stable way. For instance, it has
been shown that by modelling the bulk as a five-dimensional manifold endowed
with some kind of non-Riemannian connection we may provide the mechanism
necessary for confinement and stabilization of the motion of particles in
the brane in a purely geometrical way \cite{Alex,Formiga}.\ \ \ \ \ \ \ 

The paper is organized as follows. In Sections II and III, we present the
basic mathematical facts that define the geometries of Weyl and
Riemann-Cartan, respectively. We proceed in Section IV to briefly describe
the Rubakov-Shaposhnikov model with its mechanism of fermion confinement. In
Section V, we discuss the extension of the Dirac equation to a class of
non-Riemannian geometries. In Sections VI and VII, we investigate the
confinement of fermions induced by geometrical fields, namely, a Weyl scalar
field and a torsion field. We conclude with a few remarks in Section VIII.

\section{\protect\bigskip Weyl geometry}

We can say that the geometry conceived by Weyl is a kind of simple
generalization of Riemannian geometry in the sense that in the former one
replaces the assumption that the covariant derivative of the metric tensor $%
g $ is zero by the more general condition \cite{Weyl}

\begin{equation}
\nabla _{a}g_{bc}=\sigma _{a}g_{bc}  \label{compatibility}
\end{equation}%
where $\sigma _{a}$ \ denotes the components with respect to a local
coordinate basis of a one-form field $\sigma \ $defined on $M$. This, in
fact, represents a generalization of the Riemannian condition of
compatibility between the connection $\nabla $ and $g,$ which is equivalent
to requiring the length of a vector to remain unaltered by parallel
transport \cite{Pauli}. If $\sigma $ $=d\phi ,$ where $\phi $ is a scalar
field, then we have what is called\ an \textit{integrable Weyl geometry}. A
differentiable manifold $M$ endowed with a metric $g$ and a Weyl scalar
field $\phi $ $\ $is usually referred to as a \textit{Weyl frame}. It is
interesting to note that the Weyl condition (\ref{compatibility}) remains
unchanged\ when we go to another Weyl frame $(M,\overline{g},\overline{\phi }%
)$ by performing the following simultaneous transformations in $g$ and $\phi 
$:%
\begin{equation}
\overline{g}=e^{-f}g  \label{conformal}
\end{equation}%
\begin{equation}
\overline{\phi }=\phi -df  \label{gauge}
\end{equation}%
where $f$ is a scalar function defined on $M$.

Quite analogously to Riemannian geometry, the condition (\ref{compatibility}%
) is sufficient to determine\ the Weyl connetion\ $\nabla $\ in terms of the
metric $g$\ and the Weyl one-form field $\sigma .$ Indeed, a straightforward
calculation shows that one can express the components of the affine
connection with respect to an arbitrary vector basis completely in terms of
the components of $g$ and $\sigma $:%
\begin{equation}
\Gamma _{bc}^{a}=\{_{bc}^{a}\}-\frac{1}{2}g^{ad}[g_{db}\sigma
_{c}+g_{dc}\sigma _{b}-g_{bc}\sigma _{d}]  \label{Weylconnection}
\end{equation}%
where $\{_{bc}^{a}\}=$ $\frac{1}{2}g^{ad}[g_{db,c}+g_{dc,b}-g_{bc,d}]$
represents the Christoffel symbols, i.e. the components of the Levi-Civita
connection \footnote{%
Throughout this paper our convention is that Latin indices run from 1 to $n$%
, while Greek indices take values from $0$ to $n-1$. Capital Latin letters,
also running from $(0)$ to $(n-1)$ will be used for tetrad indices.}.

A clear geometrical insight on the properties of Weyl parallel transport is
given by the following proposition: Let $M$ be a differentiable manifold
with an affine connection $\nabla $, a metric $g$ and a Weyl field of
one-forms $\sigma $. If $\nabla $ is compatible with $g$ in the Weyl sense,\
i.e. if (\ref{compatibility}) holds, then for any smooth curve $\alpha
=\alpha (\lambda )$ and any pair of two parallel vector fields $V$ and $U$
along $\alpha ,$ we have 
\begin{equation}
\frac{d}{d\lambda }g(V,U)=\sigma (\frac{d}{d\lambda })g(V,U)
\label{covariantderivative}
\end{equation}%
where $\frac{d}{d\lambda }$ denotes the vector tangent to $\alpha $.

If we integrate the above equation along the curve $\alpha $, starting from
a point $P_{0}=\alpha (\lambda _{0}),$ then we obtain%
\begin{equation}
g(V(\lambda ),U(\lambda ))=g(V(\lambda _{0}),U(\lambda
_{0}))e^{\int_{\lambda _{0}}^{\lambda }\sigma (\frac{d}{d\rho })d\rho }
\label{integral}
\end{equation}%
Putting $U=V$ and denoting by $L(\lambda )$ the length of the vector $%
V(\lambda )$ at an arbitrary point\ $P=\alpha (\lambda )$\ of the curve,
then it is easy to see that in a local coordinate system $\left\{
x^{a}\right\} $ the equation (\ref{covariantderivative}) reduces to 
\begin{equation*}
\frac{dL}{d\lambda }=\frac{\sigma _{a}}{2}\frac{dx^{a}}{d\lambda }L
\end{equation*}

Consider the set of all closed curves $\alpha :[a,b]\in R\rightarrow M$,
i.e, with $\alpha (a)=\alpha (b).$ Then, \ we have the equation 
\begin{equation*}
g(V(b),U(b))=g(V(a),U(a))e^{\int_{a}^{b}\sigma (\frac{d}{d\lambda })d\lambda
}.
\end{equation*}
It follows from Stokes' theorem that if $\sigma $ is an exact form, that is,
\ if there exists a scalar function $\sigma $, such that $\sigma =d\phi $,
then

\begin{equation*}
\oint \sigma (\frac{d}{d\lambda })d\lambda =0
\end{equation*}%
for any loop. In other words, in this case the integral $e^{\int_{\lambda
_{0}}^{\lambda }\sigma (\frac{d}{d\rho })d\rho }$ does not depend on the
path.\ 

Historically, the geometrical theory developed by Weyl was used as a
framework of his gravitational theory, one of\ the first attempts to unify
gravity and electromagnetism. As is well known, although admirably
ingenious, Weyl's gravitational theory turned out to be unacceptable as a
physical theory, as was immediately realized by Einstein who raised
objections to the theory \cite{Pais,Pauli}. Einstein's argument was that in
a non-integrable\ Weyl geometry the existence of sharp spectral lines in the
presence of an electromagnetic field would not be possible since atomic
clocks would depend on their past history \cite{Pauli}. However, it has been
shown that a variant of Weyl geometries, known as Weyl integrable geometry,
does not suffer from the drawback pointed out by Einstein. Indeed, it is the
integral $I(a,b)=\int_{a}^{b}\sigma (\frac{d}{d\lambda })d\lambda $ that is
responsible for the difference between the readings of two identical atomic
clocks following different paths. Because in Weyl integrable geometry $I(a,b)
$ is not path-dependent it has attracted the attention of many cosmologists
in recent years \cite{Novello}. In the opinion of some authors, Weyl theory
\ "contains a suggestive formalism and may still have the germs of a future
fruitful theory " \cite{Bazin}.

\section{Riemann-Cartan geometry}

Another example of non-Riemannian geometry is the so-called Riemann-Cartan
geometry, which also represents one of the simplest generalizations of
Riemannian geometry. It constitutes the geometrical framework of a theory
formulated by E. Cartan \cite{Cartan} in an attempt to extend general
relativity when matter with spin is present. In spite of the limited
interest it has arisen among theoretical physicists since its conception
(perhaps due to the fact that it differs very little from general
relativity), some authors believe that the Einstein-Cartan theory can have
an important role in a future quantum theory of gravitation \cite{Trautman}.
Moreover, torsion cosmology has been investigated recently in connection
with the acceleration of the Universe \cite{Shie}. Finally, it should be
mentioned that torsion has been also considered in the context of
higher-dimensional scenarios, particularly in Kaluza-Klein theory and brane
models \cite{Shankar}.

A concept that is basic to the Riemann-Cartan geometry is that of \textit{%
torsion}, which is given by the following definition:\textbf{\ }Let $\nabla $%
\ be an affine connection defined on a manifold $M$ \ and $U,V\in T(M)$,
where $T(M)$ denotes the set of all tangent vector fields of $M$ . We define
the \textit{torsion }$T$ of $M$ as the mapping $T:T(M)\times T(M)$\ $%
\rightarrow T(M)$, such that 
\begin{equation}
T(U,V)=\nabla _{U}V-\nabla _{V}U-[U,V]\text{.}  \label{torsionless}
\end{equation}
If the torsion vanishes identically we say that the affine connection $%
\nabla $ is \textit{symmetric} (or, simply, \textit{torsionless}). A
manifold on which a nonvanishing torsion is present is called a \textit{%
Riemann-Cartan }manifold. Now, to establish a link between the affine
connection $\nabla $ and the metric $g$ we need a second definition: Let $M$
be a differentiable manifold \ endowed with an affine connection $\nabla $
and a metric tensor $g$ globally defined in$\ M$. We say that $\nabla $ is
compatible with $g$ if, for any vector fields $U,V,$ $W\in T(M)$,\ the
condition below is satisfied:

\begin{equation}
V[g(U,W)]=g(\nabla _{V}U,W)+g(U,\nabla _{V}W).  \label{W-compatible}
\end{equation}%
We now state an important result, which may\ be called the \textit{%
Levi-Civita extended theorem}: In a given differentiable manifold $M$
endowed with a metric $g$ on $M$, there exists only one affine connection $%
\nabla $ such that $\nabla $ is compatible with $g$ (\cite{Formiga}). In
other words,\ this means that the affine connection $\nabla $ is uniquely
determined from the metric $g$ and the torsion $T$ . ( In the torsionless
case, $\nabla $ is determined from $g$ alone \cite{do Carmo} ).

Now, a tensor that is naturally associated with $T$ is the \textit{torsion
tensor} $\mathcal{T}$ , defined by the mapping $\mathcal{T}$ \ $:T^{\ast
}(M)\times T(M)\times T(M)$\ $\rightarrow R$\ , such that $\mathcal{T}$ $(%
\widetilde{w},U,V)=$\ $\widetilde{w}(T(U,V))$, where$\ T^{\ast }(M)$ denotes
the set of all differentiable one-form fields on $M$ and $\widetilde{w}\in \
T^{\ast }(M)$. \ It is easy to see that the components of $\mathcal{T}$ \ in
a coordinate basis associated with a local coordinate system $\left\{
x^{a}\right\} $ are simply given in terms of the connection coefficients,
i.e. $\mathcal{T}$ $_{\;bc}^{a}=\Gamma _{\;bc}^{a}-\Gamma _{\;cb}^{a}$,
where $\Gamma _{\;bc}^{a}\equiv dx^{a}(\nabla _{\partial _{b}}\partial _{c})$
. A straightforward calculation shows that one can express the components of
the affine connection as%
\begin{equation}
\Gamma _{bc}^{a}=\{_{bc}^{a}\}+K_{\;bc}^{a}
\end{equation}%
where $\{_{bc}^{a}\}$ denotes the Christoffel symbols of second kind and $%
K_{\;bc}^{a}=-\frac{1}{2}(\mathcal{T}_{\;cb}^{a}+\mathcal{T}_{cb}^{\;\;a}+%
\mathcal{T}_{bc}^{\;\;a})$ represents the components of another tensor,
called the \textit{contorsion tensor} \footnote{%
Note that the indices appearing in \ the components of the torsion are
raised and lowered with $g^{ab}$ and $g_{ab}$, respectively.}.

Thus, we see that what basically makes the geometry discovered by Cartan
distinct from Riemannian geometry is simply the fact that in the latter the
affine connection $\nabla $ is not supposed to be symmetric. As a
consequence, the affine connection $\nabla $ is no longer a Levi-Civita
connection and for this reason affine geodesics do not coincide in general
with metric geodesics.

\section{Fermions confinement and the Rubakov-Shaposhnikov model}

In the context of theories of Kaluza-Klein type the compactness of extra
dimensions ensures that spacetime is effectively four-dimensional. On the
other hand, in higher-dimensional scenarios that postulate the existence of
non-compact dimensions the assumption that matter should somehow be confined
to a four-dimensionsal hypersurface is certainly a prerequisite for the
theory to be consistent with observation. One of the first atempts to
construct a higher-dimensional model that exhibits a confinement mechanism
was devised by Rubakov and Shaposhnikov \cite{Rubakov}. In this section we
give a brief account of this model.

Let us begin by considering a theory of one real scalar field $\varphi $
whose action is given by 
\begin{equation*}
S_{\varphi }=\int d^{4}xdl\text{ }[\frac{1}{2}\partial _{a}\varphi \partial
^{a}\varphi -V(\varphi )]\text{ ,}
\end{equation*}%
where $a=0,1,...4$ , leading to the equation 
\begin{equation}
\square \varphi +\frac{dV}{d\varphi }=0\text{ .}  \label{potential}
\end{equation}%
The scalar field potential is assumed to have the form $V(\varphi )=\lambda
(v^{2}-\varphi ^{2})^{2}$, and if $\varphi $ is assumed to depend only on
the extra coordinate $l$, then (\ref{potential}) yields the classical
solution (often referred to as a \textit{kink}) 
\begin{equation*}
\varphi (l)=v\tanh (2\lambda v)\text{.}
\end{equation*}%
As is well known, this solution describes a domain wall separating two
classical vacua of the model and has the following asymptotic behaviour 
\begin{equation*}
\varphi (l\rightarrow \pm \infty )=\pm v.
\end{equation*}

\ The basic idea in the Rubakov-Shaposhnikov model is to introduce a
fermionic field $\Psi $, in five-dimensional Minkowski spacetime, that
interacts with the scalar field $\varphi $ (Yukawa-type interaction) through
the action 
\begin{equation}
S_{\Psi }=\int d^{4}xdl\left[ i\overline{\Psi }\gamma ^{a}\partial _{a}\Psi
-h\varphi \overline{\Psi }\Psi \right] \text{ ,}  \label{action}
\end{equation}%
where $h$ is the coupling constant, and the matrices $\gamma ^{0}$, $\gamma
^{1}$, $\gamma ^{2}$ , $\gamma ^{3}$ are the standard Dirac matrices in four
dimensions, with $\gamma ^{4}=-\gamma ^{0}\gamma ^{1}\gamma ^{2}\gamma ^{3}$%
. It is worth noting that in each of the scalar field vacua, i.e. when $%
\varphi =\pm v$ the fermions acquire a mass given by $m=hv$.

\ The Dirac equation resulting from the action (\ref{action}) is 
\begin{equation}
i\gamma ^{a}\partial _{a}\Psi -h\varphi \Psi =0\text{ .}  \label{dirac}
\end{equation}%
It turns out that for $m=0$ there exists a zero mode solution, which is
given by 
\begin{equation}
\Psi _{0}=\Psi _{w}(p)e^{-\int_{0}^{l}dyh\varphi (y)}\text{ ,}
\label{zeromode}
\end{equation}%
where $\Psi _{w}(p)$ is the usual solution of the four-dimensional Weyl
equation \cite{Rubakov}. Note that for large values of $\left\vert
l\right\vert $ we have 
\begin{equation*}
\Psi _{0}\propto e^{-m\left\vert l\right\vert }.
\end{equation*}%
It is clear that the zero mode (\ref{zeromode}) is localized near the
hypersurface $\Sigma $ defined by $l=0$, and the probability of finding the
particle outside $\Sigma $ goes down exponentially. This result is valid for
massless fermions (e.g. neutrinos that satisfy the equation $i\gamma ^{\mu
}\partial _{\mu }\Psi _{w}=0$, where here $\mu =0,...,3$). Our purpose now
is to show that it is possible to construct field-theoretical models in
which fermions are localized by means of pure geometrical fields, i.e.
models in which the fermion confinement has a geometrical origin.

\section{Dirac equation in non-Riemannian geometries}

In this section we will assume that the geometry of the bulk is modelled by
a non-Riemannian five-dimensional manifold $M$ endowed with an affine
connection, which has a Weyl or Riemann-Cartan structure. We will assume, as
in Rubakov-Shaposhnikov model, the existence of five-dimensional fermions
that live in a four-dimensional hypersurface $\Sigma $ embedded in $M$.
However, instead of postulating an Yukawa interaction of the fermions with a
scalar field $\varphi $, we will seek a geometrical mechanism capable of
confining the fermions in $\Sigma $ without the intervention of any other
physical field. As far as the trapping mechanism is concerned the new
degrees of freedom coming from the non-Riemannian character of $M$ will play
the same role as $\varphi $. We thus need to write the Dirac equation in $M$
in such a way as to take into account the fact that the geometry now has a
non-Riemannian character. We will then look for solutions of the fermionic
field in that geometry that describe confinement in $\Sigma $ ,
corresponding, in a certain way, to a geometric analogue of the
Rubakov-Shaposhnikov model.

As is well known, the Dirac equation is a quantum mechanical wave equation
proposed by Dirac in 1928 that describes the dynamics of elementary spin
-1/2 particles\ in flat spacetime and that is fully consistent with the
principles of quantum mechanics and special relativity. In generalizing it
to Riemannian curved spacetime, one often uses the minimal coupling
procedure (MCP) and assumes that the covariant derivative of the Dirac
matrices is zero \cite{James2945}. This procedure also allows the Dirac
equation to be written in curvilinear coordinates and in non-inertial
frames. However, if the spacetime is non-Riemannian, we are in presence of
an ambiguity. Indeed, in this case we need to decide whether the MCP should
be applied to the Dirac Lagrangian, or directly to the field equations \cite%
{Hammondr20501}. As it happens, different choices lead to different field
equations. In this paper, we will consider, for simplicity, that the minimal
coupling of the fermionic field with the geometry is implemented directly in
the Dirac equation.

In what follows, we will make use of the moving frame formalism, in which
the components of the metric $g$, when written with respect to a vector
basis ("\textit{tetrad" basis}) $\left\{ e_{A}\right\} ,$ are given by $%
g(e_{A},e_{B})=\eta _{AB}=diag(1-1-1-1-1)$. The dual basis of $\left\{
e_{A}\right\} $\ will be denoted by $\left\{ \theta ^{A}\right\} $, whereas
the components of $e_{A}$ and $\theta ^{A}$\ in a coordinate basis will be
denoted, respectively, by $e_{A}^{\ \ \mu }$ and $e_{\ \ \mu }^{A}$. Tetrad
indices will be raised and lowered with Minkowski five-dimenional metric
tensor $\eta _{AB}=g(e_{A},e_{B})$.

Now, in the case of a non-Riemannian connection,\ the application the MCP
directly to the Dirac equation will consist of the following general
procedure: we start with the Dirac equation written in curved space \cite%
{Kiefer} and than replace the Riemannian metric connection by a
non-Riemannian affine connection. 
\begin{equation}
i\gamma ^{C}\left( \partial _{C}+A_{C}\mathbb{I}+\frac{1}{8}\omega
_{ACB}[\gamma ^{A},\gamma ^{B}]\right) \Psi -m\Psi =0,  \label{901a}
\end{equation}%
where $m$ is the fermionic mass, $A_{C}$ denote the components of an
arbitrary vector field $A$ (that can be regarded as an external
electromagnetic potential minimally coupled to $\Psi $), $\mathbb{I}$ is the
unit matrix, $\gamma ^{A}$ are the Dirac matrices, and $\omega _{\ \
CB}^{A}=\theta ^{A}(\nabla _{C}e_{B})$ gives the components of the affine
connection in a tetrad basis. Considering the rather general case of a
non-Riemannian manifold in which the affine connection may depend on both
the torsion and the Weyl field, the connection components $\omega _{ACB}$
can be written as 
\begin{equation}
\omega _{ACB}=e_{B\ \ ,C}^{\ \ \lambda }e_{A\lambda }+\{_{\alpha \mu
}^{\lambda }\}e_{A\lambda }e_{C}^{\ \ \alpha }e_{B}^{\ \ \mu }+\overset{w}{%
\Gamma }_{\ \!\!\!ACB}^{\mathrm{\raisebox{-0.1cm}{$_{}$}}}+K_{ACB}\text{ },
\label{1001a}
\end{equation}%
where 
\begin{equation}
\overset{w}{\Gamma }_{\ \!\!\!ACB}^{\mathrm{\raisebox{-0.1cm}{$_{}$}}%
}=-1/2(\sigma _{C}\eta _{AB}+\sigma _{B}\eta _{AC}-\eta _{CB}\sigma _{A})%
\text{ ,}  \label{072011a}
\end{equation}%
and 
\begin{equation}
K_{ACB}=-1/2(T_{CBA}+T_{BCA}-T_{ACB})  \label{072011b}
\end{equation}%
represents, respectively, the non-metric contributions of the Weyl and
contorsion field to the affine connection. Here, the components of the
torsion tensor are given by $T_{\ \ BC}^{A}=\theta ^{A}(\mathcal{T}%
(e_{B},e_{C}))$.

\section{Fermion confinement induced by a Weyl scalar field}

\label{c1001} Working along the same lines of the Rubakov-Shaposhnikov model 
\cite{Rubakov}, i.e. considering a five-dimensional torsionless Minkowski
space-time $M$ and the same convention for the Dirac matrices in four
dimensions, as mentioned above, we will also make the additional assumption
that $M$ is a Weyl integrable space. In Cartesian coordinates are $g_{\mu
\nu }=\eta _{\mu \nu }$, so we choose $e_{A}^{\ \ \mu }=\delta _{A}^{\ \ \mu
}$. In addition, in these coordinates we also have $\{_{\alpha \mu
}^{\lambda }\}=0$. Setting $K_{\mu \nu \alpha }=0$ the general expression (%
\ref{1001a}) for the affine connection $\omega _{ACB}$ reduces to $\omega
_{ACB}=$ $\overset{w}{\Gamma }_{\ \!\!\!ACB}^{\mathrm{%
\raisebox{-0.1cm}{$_{}$}}}$. By substituting $\omega _{ACB}$ into Eq. (\ref%
{901a}), it is not difficult to see that we finally get 
\begin{equation}
i\gamma ^{\mu }\partial _{\mu }\Psi -i\sigma _{\mu }\gamma ^{\mu }\Psi
-m\Psi =0,  \label{1001b}
\end{equation}%
where we have set $A=0$ as we are not considering electromagnetic
interaction, and $\sigma _{\mu }$ denotes the components of $\sigma =d\phi $
with respect to the chosen basis.

We now need to solve the Dirac equation (\ref{1001b}). If we define the new
variable $\psi $ by writing $\Psi =e^{\phi (x^{\alpha })}\psi $, then (\ref%
{1001b}) can be put in the simpler form 
\begin{equation}
i\gamma ^{\mu }\partial _{\mu }\psi -m\psi =0.  \label{1001c}
\end{equation}%
This equation is formally identical to the Dirac equation in
five-dimensional Minkowski spacetime for a free particle, whose solutions
are given, in standard notation, by 
\begin{equation}
\psi =u_{p}e^{-ip_{\alpha }x^{\alpha }},  \label{dirac solutions}
\end{equation}%
where the spinors $u_{p}$ can be normalized in a Lorentz-covariant way,
similarly to the four-dimensional case \cite{Bjorken}.\ Because the wave
functions (\ref{dirac solutions}) do not diverge as $l$ goes to $\pm \infty $%
, it is easy to choose a particular functional form of $\sigma $ such that
particle confinement takes place for any given hypersurface $\Sigma $
described by $l=constant$. As an example, let us take $\phi =-l^{2}$. With
this choice we have $\overline{\Psi }\Psi =e^{-2l^{2}}\overline{\psi }\psi $%
, which clearly means that the Weyl scalar field $\phi $ acts effectivelly
as a mechanism for localizing the fermions in the hypersurface $\Sigma $.

It is not difficult to generalize the above procedure to the case in which
the manifold $M$ , i.e the bulk, is $n$-dimensional. Indeed, in $n$
dimensions Eq. (\ref{901a}) \ reads 
\begin{equation}
i\gamma ^{\alpha }\left( \partial _{\alpha }+\Gamma _{\alpha }-\frac{n-1}{4}%
\sigma _{\alpha }\right) \Psi -m\Psi =0\text{ },  \label{general}
\end{equation}%
where $\Gamma _{\alpha }=\frac{1}{8}\left( e_{B\ \ ,\alpha }^{\ \ \lambda
}[\gamma _{\lambda },\gamma ^{B}]+\{_{\alpha \mu }^{\lambda }\}[\gamma
_{\lambda },\gamma ^{\mu }]\right) $ and, again, we are assuming that both $%
A_{C}$ and $K_{ABC}$ vanish. As in the previous case, the above equation can
be written in a simpler form if we now define the new variable $\Psi
=e^{(n-1)\phi /4}\psi $. In this way we obtain \ 
\begin{equation}
i\gamma ^{\alpha }\left( \partial _{\alpha }+\Gamma _{\alpha \ }\right) \psi
-m\psi =0\text{ }.  \label{curved dirac}
\end{equation}%
Note that the above equation is formally identical to the Dirac equation in
a Riemannian spacetime. Exact solutions of (\ref{curved dirac}) in four
dimensions that are normalizable are numerous \cite{villalba197} \ On the
other hand, in the case where $n>4$ if $\overline{\psi }\psi $ does not
diverge at $l=$ $\pm \infty $, then it is not difficult to devise a Weyl
scalar field $\phi $ such that $\overline{\Psi }\Psi =e^{(n-1)\phi /2}%
\overline{\psi }\psi $ tends to zero at large values of the extra dimension $%
l$.

\section{Fermion confinement of particles induced by torsion}

Our second example of fermion confinement induced by geometry makes use of a
torsion field. Thus, instead of postulating the existence of a Weyl field,
we will now assume that the bulk is modeled by a n-dimensional
Riemann-Cartan manifold $M$ endowed with a non-symmetric affine connection $%
\nabla $, thus giving rise to a torsion field. We then turn to the general
form of Dirac equation (\ref{901a}) and set $\sigma _{A}=0$. In addition, we
choose the torsion tensor as given by the special form 
\begin{equation}
T_{\ \ \beta \lambda }^{\alpha }=f_{\nu }(x)\left( \delta _{\beta }^{\alpha
}\delta _{\lambda }^{{\nu }}-\delta _{\lambda }^{\alpha }\delta _{\beta }^{{%
\nu }}\right) \text{ ,}  \label{torsion special}
\end{equation}%
where $f=f(x)$ is an arbitrary function to be determined later and\ $f_{\nu
} $ stands for $\partial f/\partial x^{\nu }$. Writing (\ref{072011b}) in a
coordinate basis and taking $T_{\ \ \beta \lambda }^{\alpha }$ as in (\ref%
{torsion special}) , we get $K_{\alpha \mu \beta }=f_{\nu }(g_{\alpha \mu
}\delta _{\beta }^{\nu }-g_{\beta \mu }\delta _{\alpha }^{\nu })$. It is not
difficult to verify that Eq. (\ref{901a}) reduces to 
\begin{equation}
i\gamma ^{\mu }\left( \partial _{\mu }+\Gamma _{\mu }+\frac{n-1}{2}f_{\mu
}\right) \Psi -m\Psi =0,
\end{equation}%
which, in turn, leads again to Dirac equation in a n-dimensional Riemannian
spacetime 
\begin{equation}
i\gamma ^{\mu }\left( \partial _{\mu }+\Gamma _{\mu }\right) \psi -m\psi =0%
\text{ ,}  \notag
\end{equation}%
where this time we have put $\Psi =\psi (x)e^{-(n-1)f/2}$. Therefore, we
recover the same result obtained in Section VI with the torsion field (\ref%
{torsion special}) replacing the Weyl scalar field. In this case, it is the
torsion field that provides the\ geometrical mechanism capable of trapping
matter to the brane.

\section{Final remarks}

In this paper, we have shown, inspired by the well-known
Rubakov-Shaposhnikov five-dimensional model, how it is possible to trap
fermions to a brane through purely geometrical fields. We established this
result by allowing the bulk to be described either by a Weyl space or a
Riemann-Cartan manifold. In both cases, by considering the Dirac equation in
such non-Riemannian geometries, we have shown that with an appropriate
choice of the Weyl field or the torsion it is fairly straightforward to
construct models in which fermions are localized on a brane.We also have
shown that in both cases the confinement is independent of the energy and
the mass of the fermionic particle.

The Dirac equation in non-Riemannian geometries has been studied in
different contexts by some authors \cite{Novello1}. However, in the present
work we have not assumed \textit{a priori}\ any particular theory of gravity
which takes into account the Weyl field or torsion. In other words, the
geometry of the higher-dimensional space is not assumed to come from a
particular set of field equations or theory. In a certain way, this\ seems
to give some generality to our results, although, for the sake of
completeness, it would be desirable to work out the problem in the setting
of a concrete higher-dimensional gravitational theory, a task that we leave
for the future.

\bigskip

\bigskip

\section*{Acknowledgements}

\noindent We are indebted to I. P. Lobo for helpful discussions. C.R and J.\
B. F. would like to thank CNPq for financial support.

\end{document}